\documentclass[aps,prb,twocolumn,showpacs]{revtex4}

\usepackage{times}
\usepackage{color}
\usepackage{graphicx}
\usepackage{dcolumn}
\usepackage{amssymb}
\usepackage{amsmath}
\usepackage{amsfonts}
\usepackage{sidecap}
\usepackage{bm}%
\usepackage[colorlinks=true,linkcolor=blue,pagecolor=blue,filecolor=blue,menucolor=blue,urlcolor=blue,citecolor=blue,anchorcolor=blue]{hyperref}%

\newcommand{\vm}{\boldsymbol{m}}
\newcommand{\vM}{\boldsymbol{M}}
\newcommand{\X}{\mathcal{X}}
\newcommand{\Hp}{\mathcal{H}_\perp}

\newcommand{\eg}{\textit{e.g. }}
\newcommand{\etal}{\emph{et al.}}

\begin{document}

\title{Asymmetric Ferromagnetic Resonance, Universal Walker Breakdown, and Counterflow Domain Wall Motion in the Presence of Multiple Spin-Orbit Torques}

\author{Jacob Linder}
\email{jacob.linder@ntnu.no} \affiliation{Department of Physics,
Norwegian University of Science and Technology, N-7491 Trondheim,
Norway}

\author{Mohammad Alidoust }
\email{phymalidoust@gmail.com} \affiliation{Department of Physics,
Norwegian University of Science and Technology, N-7491 Trondheim,
Norway}

\date{\today}

\begin{abstract}
We study the motion of several types of domain wall profiles in
spin-orbit coupled magnetic nanowires and also the influence of
spin-orbit interaction on the ferromagnetic resonance of uniform
magnetic films. Whereas domain wall motion in systems without
correlations between spin-space and real-space is not sensitive to
the precise magnetization texture of the domain wall, spin-orbit
interactions break the equivalence between such textures due to the
coupling between the momentum and spin of the electrons. In
particular, we extend previous studies by fully considering not only
the field-like contribution from the spin-orbit torque, but also the
recently derived Slonczewski-like spin-orbit torque. We show that
the latter interaction affects both the domain wall velocity and the
Walker breakdown threshold non-trivially, which suggests that it
should be accounted in experimental data analysis. We find that the
presence of multiple spin-orbit torques may render the Walker
breakdown to be universal in the sense that the threshold is
completely independent on the material-dependent Gilbert damping
$\alpha$, non-adiabaticity $\beta$, and the chirality $\sigma$ of
the domain wall. We also find that domain wall motion against the
current injection is sustained in the presence of multiple
spin-orbit torques and that the wall profile will determine the
qualitative influence of these different types of torques (\eg
field-like and Slonczewski-like). In addition, we consider a uniform ferromagnetic layer under a current bias, and find that
the resonance frequency becomes asymmetric against the current
direction in the presence of Slonczewski-like spin-orbit coupling. This is in
contrast with those cases where such an interaction is absent, where the
frequency is found to be symmetric with respect to the current
direction. This finding shows that spin-orbit interactions may offer additional control over pumped and absorbed energy in a ferromagnetic resonance setup by manipulating the injected current direction.

\end{abstract}

\maketitle

\section{Introduction}
Spintronics has been a highly fertile research area especially over
the last two decades \cite{zutic_rmp_04}, giving rise to practical
developments such as read-heads of harddrives, non-volatile magnetic
memory, and other types of magnetic sensors
\cite{chappert_naturemat_07, katine_jmmm_08}. The key ingredient in
this field is to utilize the spin-degree of freedom in currents and
materials to achieve the desired functionality, in particular with
an eye to providing a feasible alternative to semiconductor
technology. One of the main obstacles to overcome in this regard is
the high energy cost associated with \eg Joule heating when passing
a spin-polarized current consisting of electrons through a device:
current-densities of order 10$^6$ A/cm$^2$ are needed to perform
magnetization switching via current-induced spin-transfer torque. As
an alternative mechanism to spin-transfer torque which could
circumvent the Joule heating from electrons, magnon-induced
magnetization dynamics has been investigated more recently
\cite{han_apl_09, jamali_apl_10, yan_prl_11, linder_prb_12}.

\begin{figure}[t!]
\includegraphics[width=8.5cm,height=5.5cm]{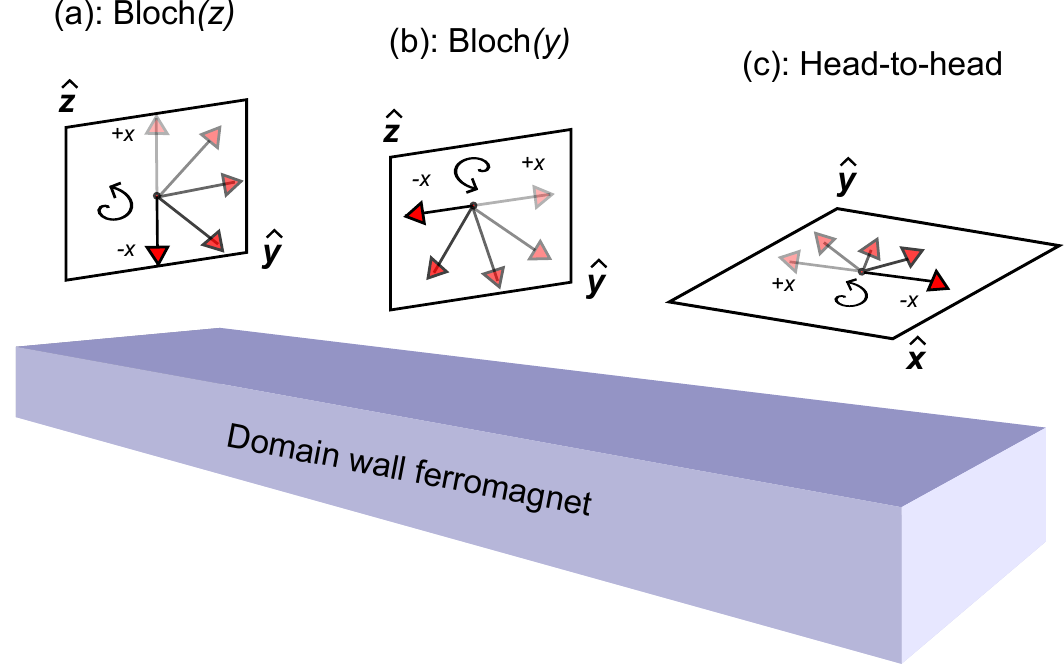}
\caption{\label{fig:model} (Color online) Schematic setup: a
spin-polarized current is passed through a domain wall magnetic
nanowire with spin-orbit coupling. The spin-orbit interaction may be
either intrinsic or induced via a heavy metal proximate host. We
consider several types of domain wall configurations, since the
presence of spin-orbit coupling qualitatively distinguishes the
domain wall motion with one type of magnetization texture from
another. More specifically, we consider two types of Bloch-domain
walls relevant for perpendicular magnetic anisotropy systems in
addition to a head-to-head domain wall with an in-plane
magnetization easy anisotropy. }
\end{figure}

Currently, the topic of controllable domain wall motion is receiving
much attention (see \eg Ref. \onlinecite{grollier_phys_11} for a
very recent review) due to its potential with regard to the storage
and transfer of information. A domain wall is a topological defect
in a magnetic system where the local magnetic order parameter
typically rotates spatially in a fashion that reduces the net
magnetic moment of the domain wall area. Owing to their small size
($\sim$ 10 nm) and large velocities ($\sim$ 100 m/s)
\cite{hayashi_prl_07, pizzini_ape_09}, controllable domain wall
motion represents holds real potential for tailoring functional
devices with fast writing speeds. In addition, there has been
several proposals \cite{kent_apl_04, matsunaga_ape_08,
parkin_science_08, liu_apl_10} related to magnetic memory
functionality due to the non-volatile nature of magnetic domains.
Walker breakdown \cite{walker} is nevertheless a limiting factor in
this regard.

Domain walls can come in several different shapes depending on the
anisotropy energies and dimensionality of the system at hand. In a
low-dimensional system such as a magnetic nanowire, Bloch walls are
one of the most frequent types encountered. However, it is also
possible to generate other sorts of magnetization textures such as
head-to-head domain walls. Both of these wall types are shown in
Fig. \ref{fig:model}. A key question is whether or not specific
domain wall types are beneficial with regard to the objectives
mentioned above (\eg fast propagation, low current densities to
generate motion). The answer to this question depends on if the spin
and position degrees of freedom are correlated in the system, for
instance via spin-orbit interaction. In the absence of such
spin-orbit interactions, different types of domain walls behave in
the same way - the exact magnetization texture has no effect and one
obtains for instance the same terminal domain wall velocity in all
cases. The fact changes when spin-orbit coupling is present since
the electron transport and spin torque now directly depends on the
precise magnetization texture, which warrants a specific study for
how domain wall motion is manifested for different types of domain
walls. A numerical investigation of this issue was recently put
forth in Ref. \onlinecite{fert_arxiv_12}.

The influence of spin-orbit coupling on domain wall motion has
recently been considered extensively in several theoretical works
\cite{obata_prb_08, manchon_prb_09, haney_prl_10, ryu_jmm_12,
kim_prb_12, linder_prb_13, fert_arxiv_12, bijl_arxiv_12}. On the
experimental stage \cite{chernyshov_naturephys_09, moore_apl_09,
kim_ape_10, miron_nature_mat_11}, it has been demonstrated that the
presence of spin-orbit coupling indeed influences the domain wall
dynamics in a non-trivial way including anomalous behavior such as
strongly enhanced domain wall velocities and induced wall motion in
the opposite direction of the electron flow. In order to explain
these findings, it was shown in Ref. \onlinecite{kim_prb_12} that
the presence of spin-orbit coupling would generate not only a
field-like torque but \textit{also} a so-called Slonczewski-like
torque \cite{slon}, named such due to its formal resemblence to
standard current-induced torques in the absence of spin-orbit
coupling. Alternatively, these two types of spin-orbit torques may
be characterized as out-of-plane and in-plane components of the
total Rashba torque \cite{wang_arxiv_11}.

Motivated by this, we will in this paper derive exact analytical
expressions for the domain wall velocity and Walker breakdown
threshold for several types of domain wall configurations when
including both types of spin-orbit torques in order to investigate
how the Slonczewski-like torque influences the physics at hand. This
way, we expand previous literature \cite{obata_prb_08} which has
only considered the field-like term and show that the inclusion of
the Slonczewski-like torque has profound impact on the domain wall
velocity and the threshold value of Walker breakdown. In fact, we
will show that the existence of this torque renders the threshold
value to be universal in the sense that it is independent on both
the Gilbert damping $\alpha$, the non-adiabiticity parameter
$\beta$, and the chirality $\sigma$ of the domain wall.

We will present a detailed derivation of the equations
of motion where possible and show precisely in which manner the
spin-orbit coupling influences both the domain wall velocity and the
Walker breakdown threshold value. Our analytical expressions show the precise conditions required
to realize domain wall motion against the current flow, as has been
experimentally observed recently \cite{miron_nature_mat_11},
and in particular how the domain wall chirality affects this phenomenon.

Finally, we investigate how the ferromagnetic resonance response of
a material (or equivalently the dissipation and pumping of energy)
is altered due to the above mentioned spin-orbit torques. 
The ferromagnetic resonance experiment is an important technique for obtaining information about 
anisotropy, magnetic damping and
magnetization reversal \cite{fmr1,fmr2,fmr3,fmr4,chen1,chen2,hasty}.
The influence of spin-polarized current on Gilbert damping and
ferromagnetic resonance have been extensively investigated in
different situations
\cite{fmr5,fmr6,fmr7,fmr8,fmr9,landeros_prb_08,landeros_prb_10}.

Considering a ferromagnetic resonance setup in the presence of a
current bias, we analytically show that the spin-orbit interactions
render the resonance frequency to become asymmetric with respect to
the direction of current injection. This is different from previous
works considering a ferromagnetic resonance setup in the presence of
spin-transfer torques, albeit without spin-orbit coupling, where the
frequency was found to be symmetric with respect to the current
direction \cite{landeros_prb_10,fmr9}.

This paper is organized as follows. In Sec. \ref{sec:theory}, we
outline the theoretical framework to be used in our analysis, namely
the Landau-Lifshitz-Gilbert (LLG) equation augmented to include the
role of spin-orbit coupling combined with a collective-coordinate
description of the domain wall. We then present our main findings in
Sec. \ref{sec:results}, in four subsections, where the LLG equation
is solved in order to obtain both the domain wall velocity, the
Walker breakdown threshold and the ferromagnetic resonance
frequency. In Subsec. \ref{sec:z} we consider Block($z$) wall
profile, in Subsec \ref{sec:y} Block($z$) wall profile is studied,
in Subsec. \ref{sec:h2h} a head-to-head domain wall structure is
investigated, and in Subsec. \ref{sec:fmr} we present and discuss
the results of absorbed power by a ferromagnetic film under current
injection in the presence of Slonczewski-like spin-orbit
interaction. We finally summarize our results and findings in Sec.
\ref{sec:summary}.

\section{Theory}\label{sec:theory}

The starting point of our analysis is the spatio-temporal
Landau-Lifshitz-Gilbert equation \cite{llg}, augmented to include
the contribution from torque terms arising due to the presence of
spin-orbit coupling. When a current-bias is applied along $x$ axis,
the full LLG equation takes the form \cite{thiaville_epl_05,
kim_prb_12}
\begin{eqnarray}\label{eq:llg}
\partial_t \vM &= -\gamma \vM \times (\boldsymbol{H}_\text{eff} + \boldsymbol{H}^{so} - \frac{\beta}{M_0} \vM\times \boldsymbol{H}^{so})\notag\\
& +\frac{\alpha}{M_0} \vM\times \partial_t\vM + \Gamma \partial_x\vM
- \frac{\beta\Gamma}{M_0}\vM\times\partial_x\vM.
\end{eqnarray}
The above equation describes the time-dynamics of the local magnetic
order parameter $\boldsymbol{M}(x,t)$. The effective field
$\boldsymbol{H}_\text{eff}$ is formally obtained by a functional
derivative of the free energy with respect to the magnetization and
will vary depending on \eg the anisotropy configuration of the wire \cite{tatara_physrep_08}.
The influence of spin-orbit interaction is captured as
an effective field:
\begin{align}
\boldsymbol{H}^{so} = \frac{\alpha_R m_e S}{\hbar e M_0(1+\beta^2)}
\hat{\boldsymbol{z}} \times \boldsymbol{j},
\end{align}
where inversion symmetry is broken in the $z$ direction and
$\alpha_R$ characterizes the strength of the spin-orbit coupling.
$S$ and $j$ is the polarization and density of the injected current,
whereas $m_e$ and $M_0$ is the electron mass and magnitude of the
magnetization, respectively. The parameter $\beta$ is known as the
non-adiabaticity parameter in the literature, a convention we shall
stick to although this terminology is not ideal \cite{xiao_prb_06}.

The terms in Eq. (\ref{eq:llg}) have the following physical
interpretation. The effective field causes a precession of the
magnetization vector $\vM$ and has two extra contributions in terms
of $\boldsymbol{H}^{so}$ and $\vM\times\boldsymbol{H}^{so}$ in the
presence of spin-orbit coupling. The former of these has the exact
form of an effective field-like torque whereas the latter has the
form of a Slonczewksi-like torque. Interestingly, this term was
conjectured to exist in the experiment of Miron
\etal~\cite{miron_nature_mat_11} in order to explain the results,
but it was only recently theoretically derived in Refs.
\onlinecite{kim_prb_12}, \onlinecite{wang_arxiv_11}. A key
observation is that the Slonczewski like spin-orbit torque depends
on the non-adiabaticity parameter $\beta$ which also appears for the
conventional non-adiabatic spin-transfer torque [last term in Eq.
(\ref{eq:llg})] as is well-known. The term $\propto \partial_x\vM$
is the adiabatic spin-transfer torque originating from the
assumption that the spin of the conduction electrons follow the
domain wall profile perfectly without any loss or spin scattering
and $\Gamma = \mu_BP/e M_0(1+\beta^2)$.

One of the main goal in this work is to compute the domain wall
velocity and analyze Walker breakdown for a domain wall nanowire
with spin-orbit coupling, considering several types of
experimentally relevant domain walls, both with in-plane and
perpendicular magnetization relative the extension of the wire
\cite{moore_apl_09, kim_ape_10, miron_nature_mat_11, fert_arxiv_12}.
We will take into account both the field-like and the
Slonczewski-like spin-orbit induced torques. We underline again that
the various magnetization textures considered in this paper will
give qualitatively different behavior for the wall velocity and
Walker threshold values precisely due to the spin-orbit interaction
which correlates spin- and real-space. For a Bloch($y$) domain wall
(see Fig. \ref{fig:model}), an exact analytical solution for the
domain wall velocity $v_\text{DW}$ is permissible and we will derive
this result in detail. For other types of domain walls, a general
expression for $v_\text{DW}$ is not possible to obtain analytically,
thus for completeness, we revert to a numerical study for these
cases. However, it is still possible to investigate analytically the
Walker breakdown threshold for these domain walls and we show that
the chirality of the domain wall conspires with the presence of
spin-orbit coupling to qualitatively alter the behavior of Walker
breakdown in spin-orbit coupled nanowires.

\section{Results and Discussion}\label{sec:results}
We shall start by investigating domain wall motion in the presence
of multiple spin-orbit torques and consider three types of domain
wall structures as shown in Fig. \ref{fig:model}. For each case, we
will focus on the domain wall velocity and the Walker breakdown
threshold value, giving exact analytical results where possible. We
note that such an exact solution for $v_\text{DW}$ constitutes the
most general analytical expression for the domain wall velocity up
to now, including fully the influence of spin-orbit coupling. We
then study the ferromagnetic resonance response of a magnetic layer
with a Slonczewski-like spin-orbit interaction with an injected
current into the plane of the layer and using the absorbed power by
the film, we drive the ferromagnetic resonance expression
analytically.

\subsection{Bloch($z$) wall}\label{sec:z}
Consider first a domain wall profile relevant for magnetic
nanowires with perpendicular anisotropy \cite{moore_apl_09,
kim_ape_10, miron_nature_mat_11} (\eg Co/Ni multilayers), namely a
so-called Bloch($z$) wall which is parametrized as:
\begin{align}\label{eq:mz}
\vm = (\sin\theta\sin\phi,\sin\theta\cos\phi,\sigma\cos\theta),
\end{align}
and a corresponding effective field:
\begin{align}\label{eq:heffz}
\boldsymbol{H}_\text{eff} &= \frac{2
A_\text{ex}}{M_0^2}\nabla^2\boldsymbol{m} - H_\perp m_x \hat{x} +
H_km_z\hat{z} + \boldsymbol{H}_\text{ext}.
\end{align}
Here, $H_\perp$ and $H_k$ are the anisotropy fields along the hard and easy axes of magnetization, respectively, whereas $\boldsymbol{H}_\text{ext}$ is an externally applied magnetic field. The parameter $\sigma=\pm1$
characterizes the chirality of the domain wall: both signs of
$\sigma$ give allowed equilibrium solutions $(\phi=0)$ of the
LLG-equation and describes a spin texture changing from positive to
negative depending on which direction one is moving in. Note that
$\sigma$ is also denoted the topological charge of the domain wall
\cite{tatara_physrep_08}: the winding direction of the local
magnetization dictates the effective "charge" since the sign of
$\sigma$ will determine the direction in which an external magnetic
field moves the domain wall. The components of the magnetization
vector depend on both space and time according to \cite{walker}
\begin{align}\label{eq:theta}
\cos\theta = \tanh\Big(\frac{x-X(t)}{\lambda}\Big),\notag\\
\sin\theta = \text{sech}\Big(\frac{x-X(t)}{\lambda}\Big).
\end{align}
Eq. (\ref{eq:theta}) is obtained by inserting the magnetization
profile $\vm$ into the LLG equation and solving for $\theta$ and
$\phi$ under equilibrium conditions (in which case $X(t)$ is a
constant and $\phi=0$). The tilt angle $\phi=\phi(t)$ is in general,
however, time-dependent and causes the domain wall to acquire a
finite component along the hard magnetization axis in an
non-equilibrium situation. A collective-coordinate description of
the domain wall motion is obtained if one may identify the
time-dependence of the domain-wall center position $X(t)$ and the
tilt angle $\phi(t)$. In general, other modes of deformation can be
allowed \cite{tatara_physrep_08}.
However, it can be shown that the domain wall
may be treated as rigid [only depending on $X(t)$ and $\phi(t)$] in
a collective-coordinate framework when the easy axis anisotropy
energy $K$ is assumed larger than its hard axis equivalent $K_\perp$
\cite{tatara_jpsj_08}, i.e. $|K|\gg|K_\perp|$.

It is useful to write down an explicitly normalized form of the
LLG-equation which we will use for all the domain wall profiles
considered in this work. We normalize all quantities to a
dimensionless form as defined by the following LLG equation:
\begin{align}\label{eq:normz}
\partial_\tau \vm &= - \boldsymbol{m}\times(\boldsymbol{\mathcal{H}}_\text{eff} +
\boldsymbol{\mathcal{H}}^{so} - \beta\boldsymbol{m}\times\boldsymbol{\mathcal{H}}^{so})\notag\\
&+\alpha\boldsymbol{m}\times\partial_\tau \vm + u
\partial_{\tilde{x}}\vm - \beta u \vm \times
\partial_{\tilde{x}}\vm.
\end{align}
In the specific case of a Bloch($z$) wall, we then have the
normalized effective field:
\begin{align}
\boldsymbol{\mathcal{H}}_\text{eff} &= 2\mathcal{A} \tilde{\nabla}^2 \vm - \mathcal{H}_\perp m_x\hat{x} + \mathcal{H}_km_z\hat{z},\notag\\
\boldsymbol{\mathcal{H}}^{so} &= \tilde{\alpha}_R u \hat{y}.
\end{align}
Inserting Eq. (\ref{eq:theta}) into Eq. (\ref{eq:normz}) leads to
one pair of equations of motion for the collective coordinates $X$
and $\phi$. These equations may be simplified by using Thiele's
approach \cite{thiele} where one integrates over $x$ and utilizes
$\int^\infty_{-\infty} \sin^2\theta \text{d}x = 2\lambda$ and
$\int^\infty_{-\infty} \sin\theta \text{d}x = \lambda\pi$. We then
find the following dimensionless equations:
\begin{equation}\label{eq:z1}
\left\{\begin{array}{c}
  \alpha\partial_{\tau}\phi - \sigma\partial_{\tau}\X = \sigma u - \frac{1}{2} \mathcal{H}_\perp \sin2\phi - \frac{1}{2} \tilde{\alpha}_R\pi u\sin\phi,\notag \\
 \partial_{\tau}\phi + \alpha\sigma\partial_{\tau}\X = \frac{1}{2}\beta
\tilde{\alpha}_R u\pi\sin\phi - \beta u\sigma
\end{array}\right..
\end{equation}
Here, $\X = X/\lambda$ is the normalized spatial coordinate of the
domain-wall center and $\tilde{\alpha}_R$ is a dimensionless measure of the strength of the spin-orbit interaction. In the
limiting case of an absent Slonczewski-like spin-orbit torque where
the terms proportional to $\beta\times\tilde{\alpha}_R$ are zero,
our results are consistent with Ref. \onlinecite{ryu_jmm_12}. The
$\sin\phi$ terms in Eq. (\ref{eq:z1}) makes an exact analytical
solution of the equations untractable. As we shall see, a similar
situation occurs for the head-to-head domain wall case.
Nevertheless, it is possible to make further progress in the present
case with regard to the appearance of so-called Walker breakdown
\cite{walker}. This phenomenon refers to a threshold value of the
current density for which the domain wall starts to rotate with a
time-dependent $\phi=\phi(\tau)$ rather than simply propagating with
a fixed magnetization texture, i.e. constant $\phi$. In general, it
is desirable with as large threshold value as possible for Walker
breakdown. We note in passing here that the presence of pinning
potentials and defects in the sample may also contribute to the
threshold value of the current, but we leave this issue for a future
work.

To investigate the velocity at which breakdown occurs, we combine
the equations of motion into a single equation for the tilt angle
$\phi$:
\begin{align}
\partial_{\tau}\phi &= \frac{1}{1+\alpha^2}\Big[\frac{1}{2}(\beta-\alpha)\tilde{\alpha}_R u\pi \sin\phi  - \sigma u(\beta-\alpha) \notag\\
&- \frac{1}{2}\alpha \mathcal{H}_\perp \sin 2\phi \Big].
\end{align}
There is no Walker breakdown as long as $\partial_{\tau}\phi=0$,
which holds when the tilt angle $\phi$ satisfies the equation:
\begin{align}\label{eq:wb1}
\sin 2\phi = \frac{(\beta-\alpha)u}{\alpha \mathcal{H}_\perp}
(\tilde{\alpha}_R \pi\sin\phi - 2\sigma).
\end{align}
Walker breakdown will occur at a velocity $u_c$ such that for
$u>u_c$ there is no stable solution for this equation. Now, for
$|\tilde{\alpha}_R\pi|<2$ the right hand side of Eq. (\ref{eq:wb1})
will have equal sign for its minimum and maximum value as $\phi$
varies from 0 to $2\pi$. Therefore, Walker breakdown will always
occur by increasing $u$: at some value $u_c$, the minimum value of
the right hand side of Eq. (\ref{eq:wb1}) will be larger than unity
and thus render the equation to be void of any solution. However, if
$|\tilde{\alpha}_R\pi|>2$, the minimum and maximum value of the
right hand side have \textit{opposite} signs. This means that there
must be a crossing of the 0 line at some values of $\phi$, and thus
an intersection with $\sin2\phi$. In effect, we can always find a
stable solution and there will be no Walker breakdown regardless of
the velocity $u$ when:
\begin{align}\label{eq:cond1}
\Big|\frac{\tilde{\alpha}_R\pi}{2}\Big|>1.
\end{align}

\begin{figure}[b!]
\centering \resizebox{0.42\textwidth}{!}{
\includegraphics{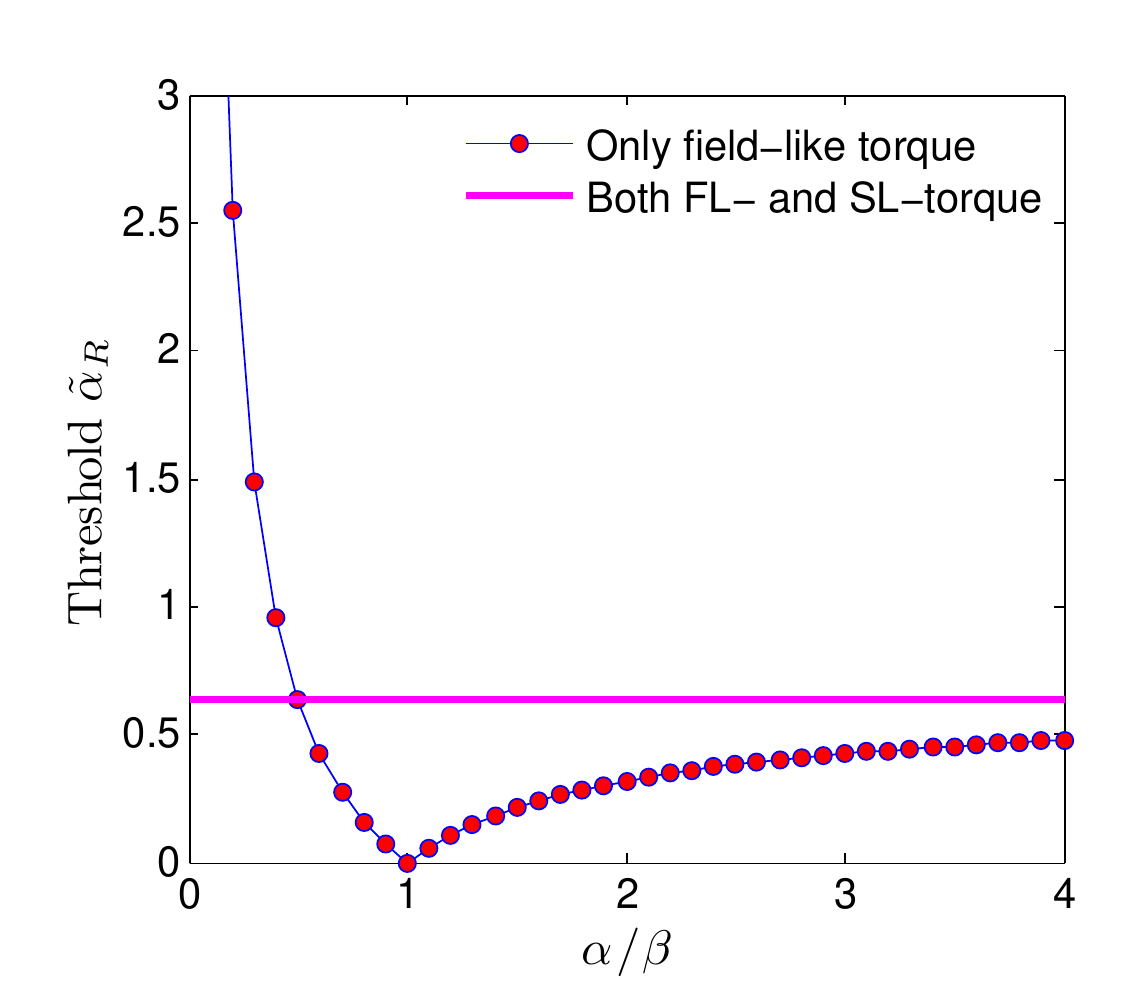}}
\caption{(Color online) Threshold value for the magnitude of the
spin-orbit coupling above which there is no Walker breakdown. In the
more general scenario where both types of spin-orbit torques are
accounted for, the threshold value for $\tilde{\alpha}_R$ is
constant. When only the field-like torque is considered, the
threshold is strongly increased in the regime $\alpha/\beta<0.5$. In
the limit $\alpha/\beta\to\infty$, the asymptote is $2/\pi$. }
\label{fig:threshold_z}
\end{figure}
In other words, for a sufficiently large spin-orbit interaction, no
Walker breakdown occurs. It is interesting to note that this
condition is universal in the sense that it is independent on the
damping parameter $\alpha$, the non-adiabiticity parameter $\beta$,
and the chirality $\sigma$ of the domain wall. This observation can
be attributed directly to the presence of the new spin-orbit torque
proportional to $\beta$. To see this, consider a scenario where only
the field-like spin-orbit torque $\propto \vM \times
\boldsymbol{H}_{so}$ is included. All terms proportional to
$\beta\times\tilde{\alpha}_R$ are then zero, and we obtain the
equation
\begin{align}
\sin2\phi = \frac{2\sigma u}{\mathcal{H}_\perp}\Big(1 - \beta/\alpha
- \frac{\sigma \tilde{\alpha}_R\pi}{2}\sin\phi\big),
\end{align}
which must be satisfied to prevent Walker breakdown. As seen,
whether or not the maximum and minimum value of the right hand side
have equal sign depends on if \begin{align}\label{eq:cond2}
\Big|\frac{\tilde{\alpha}_R\pi}{2}\Big|>|(1-\beta/\alpha)|.
\end{align}
In this regime, we recover the results of Ref.
\onlinecite{ryu_jmm_12}. The effect of the Slonczewski-like
spin-orbit torque is then to render the Walker breakdown universal
(independent on $\alpha,\beta,\sigma$). Let us also consider the
implications this torque-term has with regard to the magnitude of
the threshold value for Walker breakdown. Comparing Eqs.
(\ref{eq:cond1}) and (\ref{eq:cond2}), we see that the required
spin-orbit interaction $\tilde{\alpha}_R$ to completely remove the
Walker threshold depends on the ratio $\beta/\alpha$ if one does not
take into account the Slonczewski-like spin-orbit torque. For
$\beta/\alpha\simeq1$, the required spin-orbit strength becomes very
small. In the more general case where the aforementioned torque is
included, however, the required $\tilde{\alpha}_R$ has a fixed
value. This is shown in Fig. \ref{fig:threshold_z}.

\begin{figure}
\includegraphics[width=7.0cm,height=9.cm] {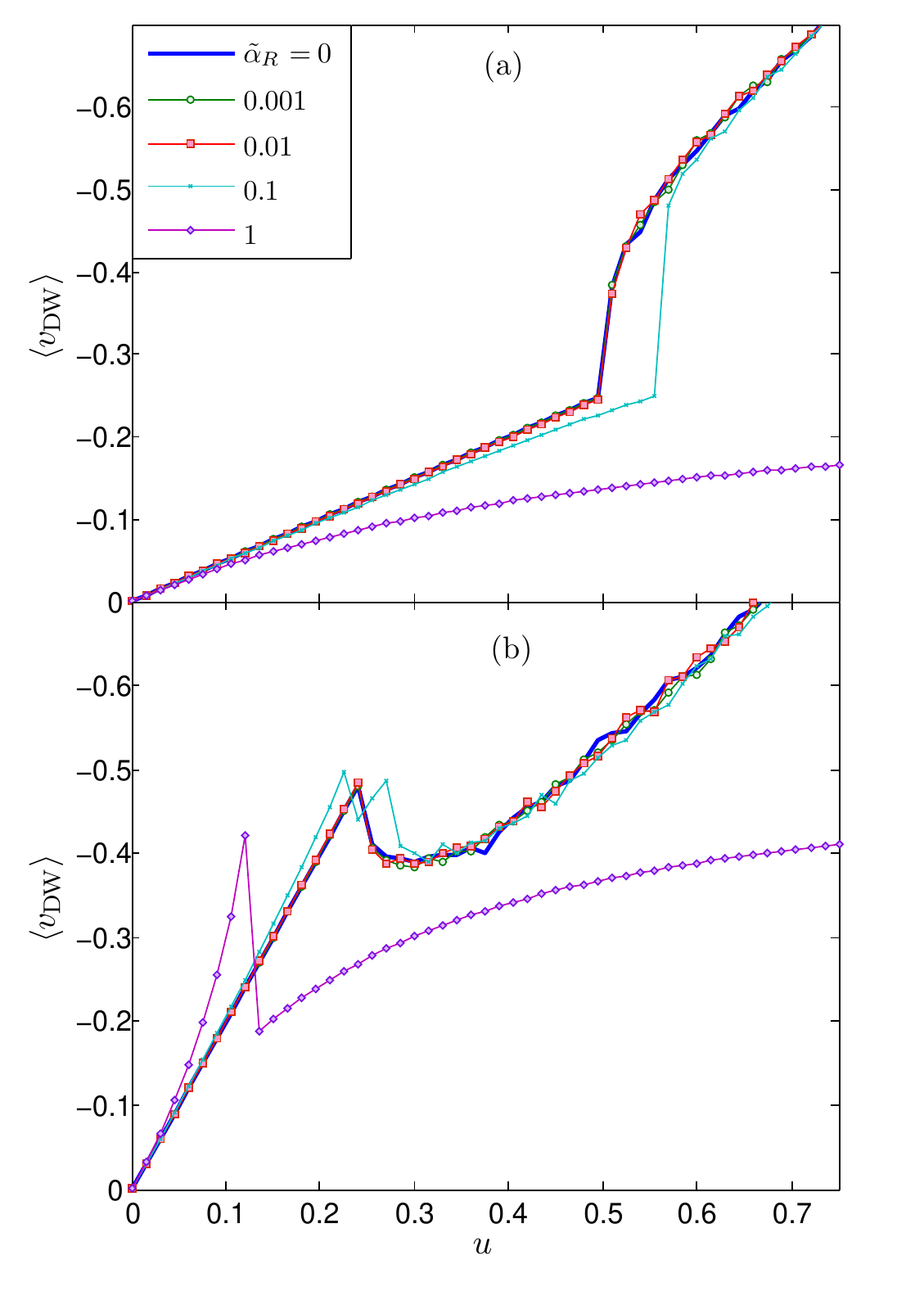}
\caption{(Color online) Domain wall velocity for a Bloch($z$) wall
plotted against the injected current. We have chosen $\sigma=1$
without loss of generality (see text) and set $\beta=0.01$ and
$\mathcal{H}_\perp$=0.5. In (a) $\alpha>\beta$ ($\alpha=0.02$)
whereas in (b) $\alpha<\beta$ ($\alpha=0.005$). Note the inverted
sign of the $y$ axis, which simply corresponds to the direction of
the wall motion. } \label{fig:wall_velocity_z}
\end{figure}

We also give numerical results for the wall velocity for this Bloch
domain wall configuration, using a similar approach as in Ref.
\onlinecite{linder_prb_11}. Let us first note that it is possible to
infer what the qualitative effect is of the chirality $\sigma$
directly from the equations of motion Eqs. (\ref{eq:z1}). By making
the transformation $\phi \to \sigma\phi$, it is seen that the
equations of motion become independent on the chirality $\sigma.$
This means that the domain wall velocity will be the same regardless
of the sign of $\sigma$, whereas the tilt angle $\phi$ evolves in
the opposite direction with time for opposite signs of $\sigma$. In
Fig. \ref{fig:wall_velocity_z}, we therefore present results for
$\sigma=1$ without loss of generality and consider two cases with
damping $\alpha$ larger or smaller than the non-adiabaticity
constant $\beta$ in (a) and (b), respectively. As seen, this
qualitatively affects the domain wall velocity.

\begin{figure*}[t!]
\centering
\resizebox{0.96\textwidth}{!}{
\includegraphics{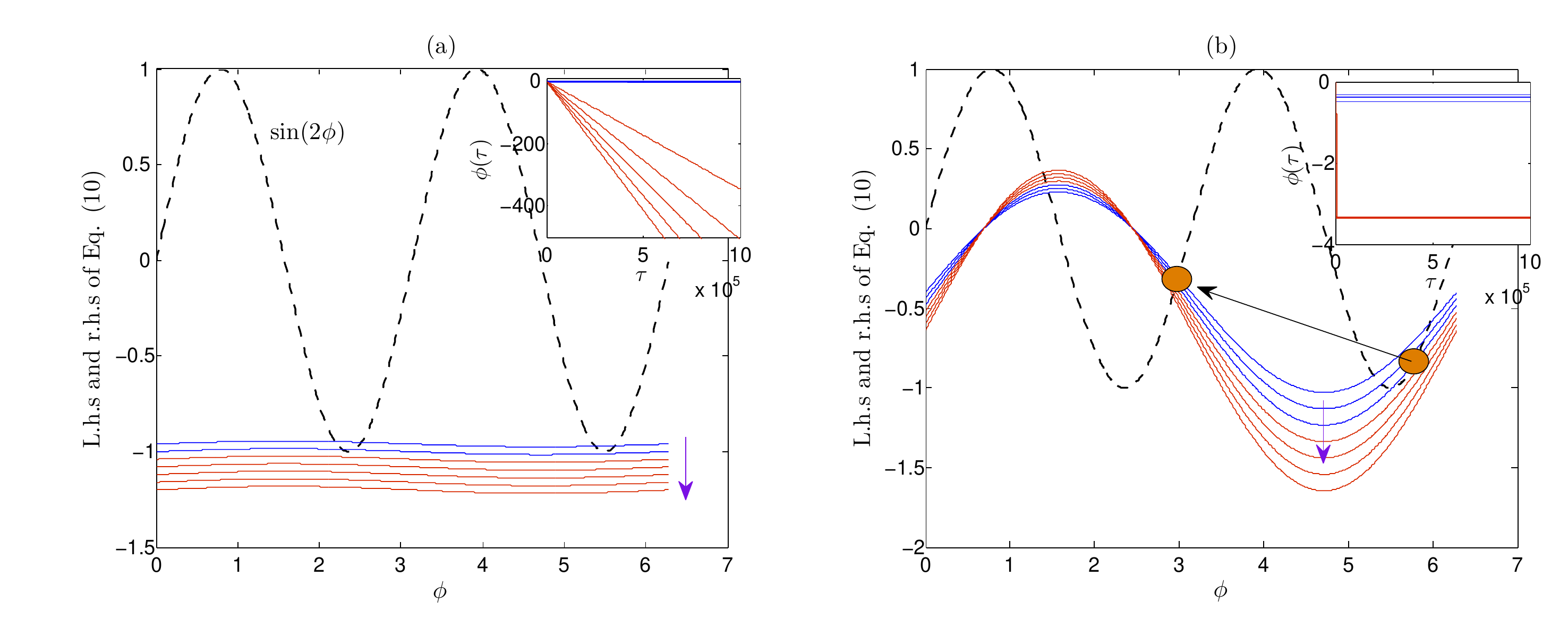}}
\caption{\label{fig:analysis} (Color online)
Plot of left-hand side (dashed line) and right-hand side (full lines) of Eq. (\ref{eq:wb1}) in order to illustrate the intersection points. When there is no intersection between the lines, Walker breakdown has occurred. We have set $\beta=0.01$, $\alpha=0.005$ and consider (a)
$\tilde{\alpha}_R=0.01$ and $u$ ranging from 0.24 to 0.30 along the direction of the arrow, in addition to (b)
$\tilde{\alpha}_R=1$ and $u$ ranging from 0.10 to 0.16 along the direction of the arrow. The black arrow between the circles in (b) highlights how the intersection point changes abruptly upon increasing $u$. \textit{Insets:} Time-evolution of the tilt angle for the same choices of $u$.}
\end{figure*}

A particular feature worth noting in (b) is that the abrupt change
in wall velocity at a given $u$ is not necessarily synonymous with
the occurrence of Walker breakdown. To see this, consider Fig. \ref{fig:analysis}
where we have plotted the left- and right-hand side of the Walker breakdown criterion
Eq. (\ref{eq:wb1}) in addition to the time-evolution of the tilt angle $\phi$ as an inset.
We have set $\alpha=0.005$ and $\beta=0.01$ and consider two strengths of the spin-orbit
coupling parameter $\tilde{\alpha}_R$ in (a) and (b). An intersection of the lines in
the main panels means that there exists a solution to Eq. (\ref{eq:wb1}) and that
Walker breakdown does not occur. Considering Fig. \ref{fig:analysis}(a) first,
we see that increasing the current density eventually causes Walker breakdown as
the dashed and full lines no longer intersect. As a result, $\phi$ is no longer a
constant as seen in the inset and starts to grow with time. We may therefore conclude
that the abrupt change in wall velocity for $\tilde{\alpha}_R=0.01$ seen in
Fig. \ref{fig:wall_velocity_z}(b) does correspond to the occurrence of Walker breakdown.
However, turning to Fig. \ref{fig:analysis}(b) it is seen that the dashed and full lines
always intersect even when increasing the current density $u$ above the value at which the
wall velocity abruptly changes in Fig. \ref{fig:wall_velocity_z}(b) for $\tilde{\alpha}_R=1$
(around $u=0.14$). What is important to note is that their point of intersection changes
discontinuously: the tilt angle $\phi$ remains constant so that there is no Walker breakdown
in the sense of a continuously deforming domain wall. Instead, there is an abrupt
change in the tilt angle where it changes from one constant value to another.

\subsection{Bloch($y$) wall}\label{sec:y}
Another type of domain wall structure which may appear in such a
systems with perpendicular magnetic anisotropy is the
Bloch$(y)$-wall, having the easy magnetization direction along the
$y$ axis whereas the hard axis remains along the wire direction:
\begin{align}
\vm = (\sin\theta\sin\phi,\sigma\cos\theta,\sin\theta\cos\phi),
\end{align}
and a corresponding effective field:
\begin{align}
\boldsymbol{H}_\text{eff} &= \frac{2
A_\text{ex}}{M_0^2}\nabla^2\boldsymbol{m} - H_\perp m_x \hat{x} +
H_km_y\hat{y} + \boldsymbol{H}_\text{ext}.
\end{align}
In this case, the equations of motion for the collective coordinates
$\X$ and $\phi$ take a different form compared to the Bloch($z$)
case:
\begin{equation}\label{eq:phiy}
    \left\{\begin{array}{c}
       \sigma\partial_{\tau}\X + \alpha\partial_{\tau}\phi = \beta\tilde{\alpha}_R u - \frac{1}{2}\mathcal{H}_\perp\sin2\phi - u\sigma,\notag \\
       \partial_{\tau}\phi - \alpha\sigma \partial_{\tau}\X = \beta u\sigma +
\tilde{\alpha}_Ru
     \end{array}\right..
\end{equation}
In fact, these equations can now be solved analytically in an exact
manner, using a similar approach as in Ref.
\onlinecite{linder_prb_13}. Combining the two above equations
yields:
\begin{align}\label{eq:phiy}
\partial_{\tau}\phi(1+\alpha^2) = -\frac{\alpha}{2}
\mathcal{H}_\perp\sin2\phi + u[\sigma(\beta-\alpha)
+\tilde{\alpha}_R(1+\alpha\beta)].
\end{align}
Consider Eq. (\ref{eq:phiy}) with respect to $\phi=\phi(\tau)$. This
is a separable equation and direct integration gives:
\begin{align}\label{eq:r1}
\tau &= \mathcal{C}_0
-\frac{1+\alpha^2}{\sqrt{\mathcal{A}^2-\alpha^2\mathcal{H}_\perp^2/4}}\text{atan}\Big[
\frac{\alpha \Hp/2 -
\mathcal{A}\tan\phi}{\sqrt{\mathcal{A}^2-\alpha^2\mathcal{H}_\perp^2/4}}
\Big],
\end{align}
where $\mathcal{C}_0$ is an integration constant and we define:
\begin{align}
\mathcal{A} \equiv u[\sigma(\beta-\alpha)
+\tilde{\alpha}_R(1+\alpha\beta)].
\end{align}
For brevity of notation, we also introduce
$\mathcal{B}\equiv\alpha\Hp/2$. The integration constant depends on
the initial conditions. At $\tau=0$, we assume that the domain wall
is in its equilibrium configuration $\phi=0$, in which case we may
write the solution for the tilt angle as:
\begin{align}
\tan\phi &= \frac{\mathcal{B}}{\mathcal{A}} - \frac{\sqrt{\mathcal{A}^2-\mathcal{B}^2}}{\mathcal{A}}\tan\Big[\text{atan}(\alpha \mathcal{B}/\sqrt{\mathcal{A}^2-\mathcal{B}^2}) \notag\\
&- \tau\sqrt{\mathcal{A}^2-\mathcal{B}^2}/(1+\alpha^2)\Big].
\end{align}
Having now obtained the full time-dependence of the tilt-angle, we
insert this back into the original equation of motion in order to
find the domain wall velocity $\dot{\X} = v_\text{DW}$. The general
expression for the domain wall velocity is rather large. However, by
utilizing the fact that $v_\text{DW}$ will display small-scale
oscillations it is possible to find a simplified expression for the
average domain wall velocity $\langle v_\text{DW} \rangle$. The
period of oscillation is $T =
(1+\alpha^2)\pi/\sqrt{\mathcal{A}^2-\mathcal{B}^2}$, which gives us:
\begin{widetext}
\begin{align}
\langle v_\text{DW} \rangle &= \frac{1}{T} \int^T_0 \text{d}\tau \frac{\sigma}{\alpha} \Bigg\{ \frac{\mathcal{A}^2-\mathcal{B}^2}{\mathcal{A}(1+\alpha^2)}\text{sec}^2\Big( \text{atan}(\alpha \mathcal{B}/\sqrt{\mathcal{A}^2-\mathcal{B}^2})-\tau\frac{\sqrt{\mathcal{A}^2-\mathcal{B}^2}}{1+\alpha^2} \Big)\notag\\
&\times \Big[1 + \Big( \frac{\alpha \mathcal{B}}{\mathcal{A}} -
\sqrt{\mathcal{A}^2-\mathcal{B}^2}{\mathcal{A}}
\tan[\text{atan}(\alpha
\mathcal{B}/\sqrt{\mathcal{A}^2-\mathcal{B}^2}) -
\tau\sqrt{\mathcal{A}^2-\mathcal{B}^2}/(1+\alpha^2)] \Big)^2
\Big]^{-1} \Bigg\} - u(\tilde{\alpha}_R\sigma + \beta)/\alpha.
\end{align}
\begin{SCfigure*}
 \centering
\includegraphics[width=14.80cm,height=8.5cm]%
{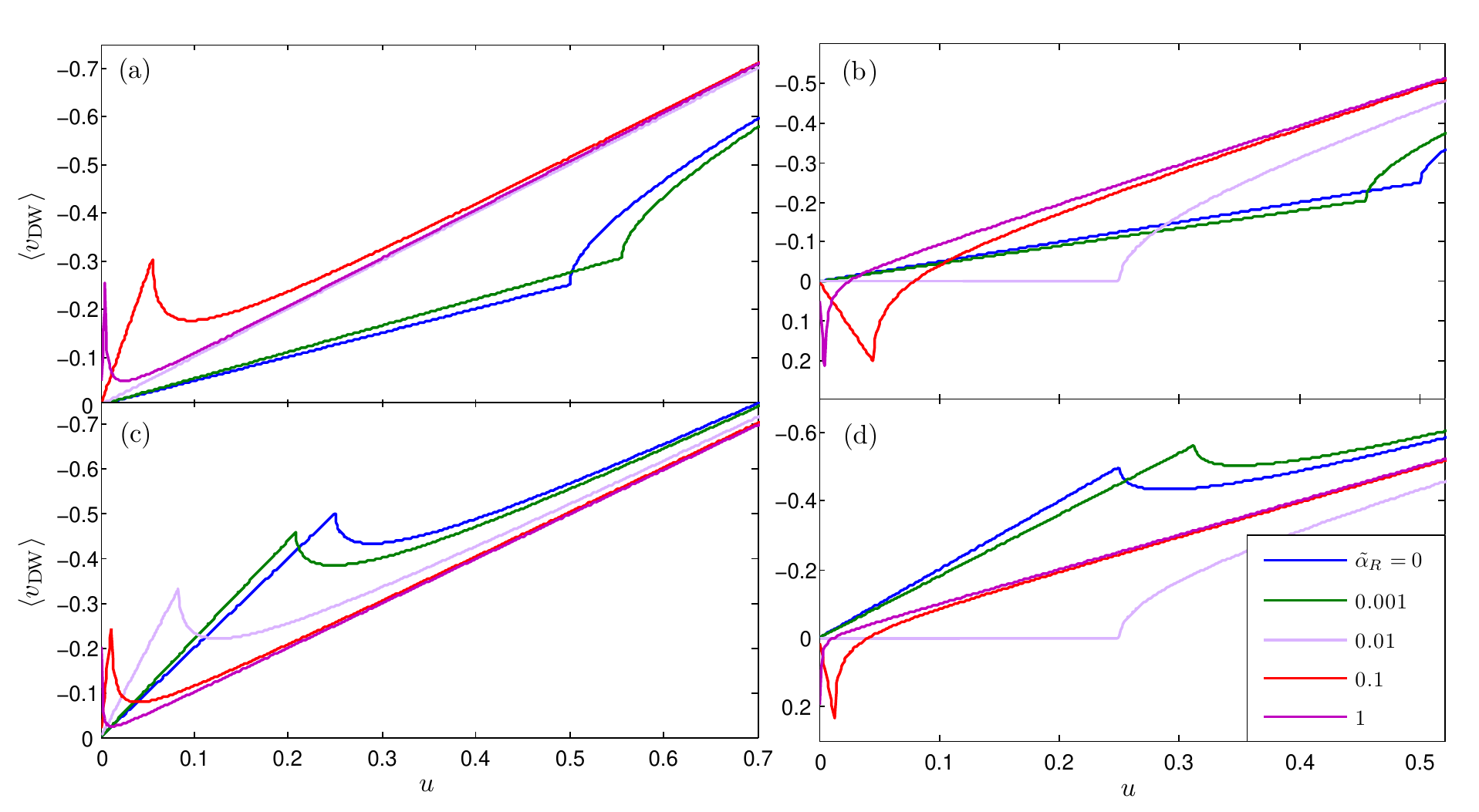} \caption{(Color online)  The domain wall
velocity $\langle v_\text{DW} \rangle$ as a function of the current
density $u$ for various chiralities and spin-orbit coupling
strengths. (a): Positive chirality $\sigma=+1$ and $\alpha>\beta$
($\alpha=0.02$). (b): Negative chirality $\sigma=-1$ and
$\alpha>\beta$ ($\alpha=0.02$). (c): Positive chirality $\sigma=+1$
and $\alpha<\beta$ ($\alpha=0.005$). (d): Negative chirality
$\sigma=-1$ and $\alpha<\beta$ ($\alpha=0.005$). For all plots, we
have used $\beta=0.01$ and $\mathcal{H}_\perp=0.5$.} \label{fig:blochy_velocity}
\end{SCfigure*}
The analytical solution to the above integral and the final result
is:
\begin{align}\label{eq:vv}
\langle v_\text{DW} \rangle &= \frac{\sigma}{\alpha(1+\alpha^2)}
\text{sgn}\{u\sigma(\beta-\alpha) +
u\tilde{\alpha}_R(1+\alpha\beta)\}\times\text{Re}\sqrt{[u\sigma(\beta-\alpha)
+ u\tilde{\alpha}_R(1+\alpha\beta)]^2 -\alpha^2\Hp^2/4}
-u(\tilde{\alpha}_R\sigma +\beta)/\alpha,
\end{align}
where we have reinstated the original parameters contained in the
quantities $\mathcal{A}$ and $\mathcal{B}$.
\end{widetext}
The equation for $\langle v_\text{DW} \rangle$ shows the exact
manner in which the domain wall velocity depends on the various
torque terms such as the non-adiabatic contribution $\beta$ and the
spin-orbit terms $\tilde{\alpha}_R$, and reveals several important
features. It is seen that for this particular domain wall
configuration [Bloch$(y)$], the effect of the Slonczewski-like
spin-orbit torque is a small quantitative correction of order
$\mathcal{O}(\alpha\beta)$, which thus can be neglected. However,
the conventional field-like spin-orbit torque has a strong
qualitative influence on the wall dynamics. In fact, it is seen that
the $\tilde{\alpha}_R$ term plays the same role as the non-adiabatic
conventional torque proportional to $\beta$, but with one important
difference: the spin-orbit torque contribution is chirality
dependent, i.e. changes sign with $\sigma$, whereas the $\beta$-term
does not. As a consequence, the wall may actually propagate in
opposite direction of the applied current depending on the chirality
$\sigma$ of the domain wall, as was shown recently in Ref.
\onlinecite{linder_prb_13}.

It is seen from Eq. (\ref{eq:vv}) that there is either an
enhancement of the domain wall velocity or a competition between the
spin-orbit induced torque and $\beta$-torque depending on the sign
of $\sigma$. We show this in Fig. \ref{fig:blochy_velocity} where we
consider the four possible combinations of wall chirality $\sigma$
(two values, $\sigma=\pm1$) combined with whether or not $\alpha$ is
larger than $\beta$ (two possibilities, $\alpha>\beta$ or
$\alpha<\beta$). For a positive chirality $\sigma=+1$ displayed in
Fig. \ref{fig:blochy_velocity} (a) and (c), the wall moves in the
same direction for all current densities $u$ as the torque terms in
Eq. (\ref{eq:vv}) have the same sign. This is no longer the case for
the opposite chirality $\sigma=-1$ shown in Fig.
\ref{fig:blochy_velocity}(b) and (d) where the wall velocity can
actually change sign as $u$ increases. This is indicative of
counterflow domain wall motion where the wall moves in the opposite
direction of the applied spin current.

Walker breakdown for the domain wall occurs for velocities $u\geq
u_c$ where the root in Eq. (\ref{eq:vv}) becomes imaginary, namely:
\begin{align}
u_c = \frac{\alpha\Hp}{|2\sigma(\beta-\alpha) +
2\tilde{\alpha}_R(1+\alpha\beta)|}.
\end{align}
Note that this is the same as $u_c$ that we would have found using
the arguments in the previous section in order to identify the
Walker breakdown from the equations of motion (without actually
solving them explicitly) and thus serves as a consistency check for
the correctness of Eq. (\ref{eq:vv}). This expression is quite
generally valid, including the effects of both types of spin-orbit
torques and both types of conventional spin-transfer torques. As
another consistency check, we observe that in the absence of
spin-orbit coupling ($\tilde{\alpha}_R = 0$), one finds that $|u_c|
= \alpha \Hp/2|\beta-\alpha|$ which agrees with Ref.
\onlinecite{thiaville_epl_05}. The effect of the spin-orbit
interaction is seen to depend explicitly on the chirality $\sigma$
of the domain wall. Although Walker breakdown is inevitable for the
present Bloch($y$) domain wall, in contrast to the Bloch($z$) one,
the presence of spin-orbit interactions $(\tilde{\alpha}_R\neq0$)
can strongly enhance the threshold velocity due to the competition
between the terms $\sigma(\beta-\alpha)$ and
$\tilde{\alpha}_R(1+\alpha\beta)$ in the denominator. When these
terms have different sign (either for $\sigma=-1$ and $\beta>\alpha$
or $\sigma=1$ and $\beta<\alpha$), the spin-orbit coupling can very
strongly enhance the threshold current for Walker breakdown. This
effect could be used to infer information about the value of
$\alpha$ and $\beta$ precisely due to the non-monotonic behavior of
the threshold current as a function of $\tilde{\alpha}_R$.

\begin{figure}
\includegraphics[width=7.cm,height=8.4cm] {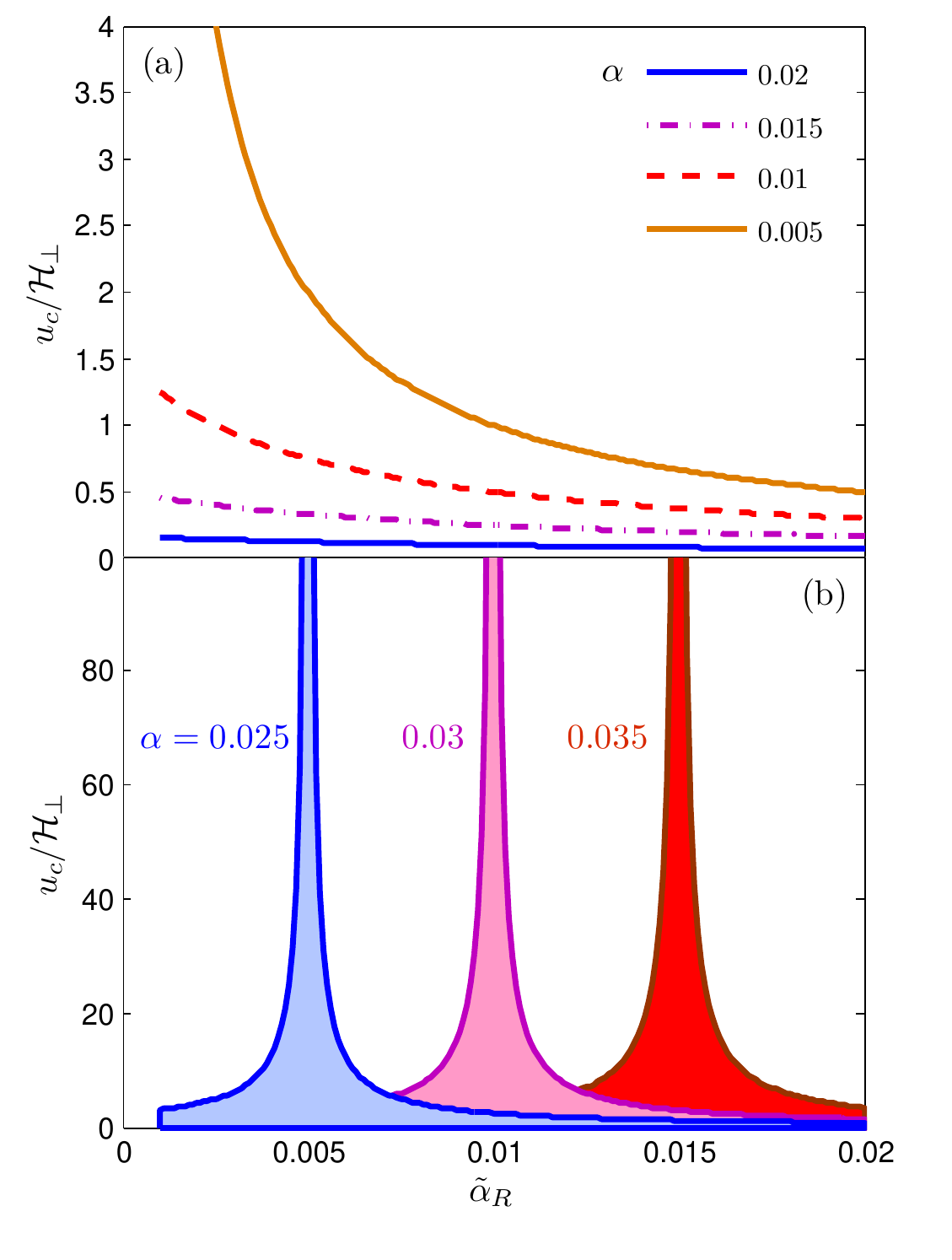}
\caption{(Color online) Critical velocity $u_c/\mathcal{H}_\perp$
that triggers Walker breakdown. We have chosen $\beta=0.02$ as a
representative value which demonstrates the fundamental behavior of
$u_c$. For sufficiently low damping $\alpha<\beta$ shown in (a), the
threshold velocity is lowered monotonically as the spin-orbit
interaction $\tilde{\alpha}_R$ is increased. When the damping
becomes stronger such that $\alpha>\beta$, $u_c$ is strongly
enhanced in a limited interval of $\tilde{\alpha}_R$.}
\label{fig:threshold_y}
\end{figure}

We illustrate this behavior in Fig. \ref{fig:threshold_y} where we
have chosen $\sigma=+1$. As seen, the threshold velocity decreases
in a monotonic fashion with increasing $\tilde{\alpha}_R$ when the
damping is low, $\alpha<\beta$. However, when the two terms in the
denominator differ in sign (which occurs precisely when
$\alpha>\beta$), the threshold velocity $u_c$ has a non-monotonic
behavior and is in fact strongly increases near $\tilde{\alpha}_R =
|\beta-\alpha|$. In this way, one may obtain information regarding
the relative size of $\alpha$ and $\beta$ by measuring the threshold
velocity.

\subsection{Head-to-head domain wall}\label{sec:h2h}
The final type of domain wall structure we will consider appears for
in-plane magnetized strips (\eg NiFe layer \cite{fert_arxiv_12}) and
is known as a so-called head-to-head domain wall. In this case, the
easy axis is parallell with the extension of the wire whereas the
hard axis is perpendicular to it:
\begin{align}
\vm = (-\sigma\cos\theta,\sin\theta\cos\phi,\sin\theta\sin\phi),
\end{align}
and a corresponding effective field:
\begin{align}
\boldsymbol{H}_\text{eff} &= \frac{2
A_\text{ex}}{M_0^2}\nabla^2\boldsymbol{m} - H_\perp m_z \hat{z} +
H_km_x\hat{x} + \boldsymbol{H}_\text{ext}.
\end{align}
Using again Thiele's approach as described in the previous sections,
one arrives at \textit{exactly} the same equations of motion as in
the Bloch($z$) case. The formal reason for this can be traced back
to the fact that the effective spin-orbit field
$\boldsymbol{H}^{so}$ is directed along the $y$ axis. The
magnetization textures of the Bloch($z$) and head-to-head domain
walls may be transformed into each other via an SO(3) rotation with
an angle $\pi/2$ of $\vM$ around the $y$ axis. Such a rotation
leaves $\boldsymbol{H}^{so}$ invariant and one thus obtains the same
equations of motion for both types of domain walls. Formally, one
can see this by multiplying Eq. (\ref{eq:llg}) from the left side
with:
\begin{align}
\mathcal{U} = \begin{pmatrix}
0 & 0 & -1\\
0 & 1 & 0\\
1 & 0 & 0 \\
\end{pmatrix},
\end{align}
and using that
\begin{align}
(\mathcal{U} \boldsymbol{a}) \times (\mathcal{U}\boldsymbol{b}) =
\text{det}(\mathcal{U}) (\mathcal{U}^{-1})^\text{T}
(\boldsymbol{a}\times\boldsymbol{b}).
\end{align}
Since $\mathcal{U}\in$ SO(3), we have that
$(\mathcal{U}^{-1})^\text{T}=\mathcal{U}$ and det$(\mathcal{U})$=+1.
By direct multiplication, one observes that
$\mathcal{U}\boldsymbol{H}^{so} = \boldsymbol{H}^{so}$, $\mathcal{U}
\vM_\text{Bloch$(z)$} = \vM_\text{head-to-head}$ and $\mathcal{U}
\boldsymbol{H}^\text{eff}_\text{Bloch$(z)$} =
\boldsymbol{H}^\text{eff}_\text{head-to-head}$. Note that it is in
drastic contrast with the Bloch($y$) case where
$\boldsymbol{H}^{so}$ is \textit{not} invariant under the matrix
which rotates $\vM_\text{Bloch$(z)$}$ into $\vM_\text{Bloch$(y)$}$.
The same arguments and results related to the domain wall velocity
and Walker breakdown that were discussed in Sec. \ref{sec:z} then
also hold for the present head-to-head domain wall case.

We mention here that the equivalence of the Bloch($z$) and
head-to-head domain wall case found here is contingent on the
specific setup we have considered in Fig. \ref{fig:model}. Although
this model is the standard one and indeed the most frequently
employed setup experimentally, it was recently shown that such an
equivalence does not hold when combining a magnetic strip/wire with
a non-magnetic conductive layer with spin-orbit interaction in a
\textit{non-parallell} geometry \cite{fert_arxiv_12}. Such a method
actually provides a manner in which the direction of the effective
spin-orbit field can be changed which could then serve as a mean to
distinguish between different types of domain walls, based on their
response to an applied current.

\subsection{Ferromagnetic resonance (FMR) in the presence of spin-orbit
torques}\label{sec:fmr}

We now turn our attention to another setup
where the aim is to identify the ferromagnetic resonance response of
a material where spin-orbit interactions play a prominent role. To
do so, we consider the setup shown in Fig. \ref{fig:model_2} where a
spin-current with polarization magnitude and unit vector direction
$S\in[0,1]$ and $\vec{\ddot{S}}$, respectively, is injected into the
ferromagnetic layer where spin-orbit coupling is present. This
directly influences the susceptibility tensor and thus both the
ferromagnetic resonance frequency/linewidth and the
absorbed power by the system \cite{prb_86_dreher}.

To facilitate the analytical calculations, we will operate with two
different coordinate systems. The laboratory (stationary) framework
$\ddot{x}\ddot{y}\ddot{z}$ is shown in Fig. \ref{fig:model_2}, where
the $\ddot{x}\ddot{y}$ plane spans the ferromagnetic layer, and
$xyz$ denotes a rotated coordinate system which we will specify the direction and purpose of below. A current is injected into the
ferromagnetic layer acting with a spin-transfer torque on the
magnetization vector $\vec{\ddot{\mathcal{M}}}$. This torque is
modified due to the presence of spin-orbit coupling which is taken
into account via a field $\vec{\ddot{H}}^{so}$ as in the domain-wall treatment. The time-dependent LLG motion equation
describing the dynamic of ferromagnetic layer magnetization vector
then takes the following form in this new notation:
\begin{eqnarray}\label{LLG}
    &&\nonumber\frac{\partial \vec{\ddot{\mathcal{M}}}}{\partial
    t}=-\gamma\vec{\ddot{\mathcal{M}}}\times\vec{\ddot{H}}^t+\frac{\alpha}{\ddot{\mathcal{M}}_{S}}\vec{\ddot{\mathcal{M}}}
    \times\frac{\partial \vec{\ddot{\mathcal{M}}}}{\partial
    t}\\&&+\frac{\gamma}{\ddot{\mathcal{M}}_S}\vec{\ddot{\mathcal{M}}}
    \times\vec{\ddot{\mathcal{M}}}\times(\beta\vec{\ddot{H}}^{so}+P_{s}
    \vec{\ddot{S}}),\\&&\nonumber \vec{\ddot{H}}^{so}=\frac{\alpha_R m_e S}{\hbar e \mathcal{\ddot{M}}_S
(1+\beta^2)}(\vec{\ddot{n}}\times\vec{\ddot{J}}_e),\;\;P_{s}=\frac{\hbar
S J_e}{2e\mathcal{\ddot{M}}_Sd}.
\end{eqnarray}
Here, $\gamma$ is the electron gyromagnetic ratio and $\alpha$ is
the Gilbert damping constant. Moreover, $\beta$ is the
non-adiabaticity parameter discussed previously, $P_s$
is the spin-torque parameter, $S$ is the polarization of injected
current into the ferromagnetic layer, and a normal vector to the
plane of ferromagnetic layer is represented by $\vec{\ddot{n}}$ (see
Fig. \ref{fig:model_2}).

We now introduce a rotated coordinate system $xyz$ where the
saturation magnetization direction is parallel with the $z$ axis.
The orientation of the rotated system $xyz$ compared to the
stationary one $\ddot{x}\ddot{y}\ddot{z}$ is determined by
calculating the equilibrium orientation of the magnetization order
parameter and setting the $z$ axis to be parallel with it. The
details of the calculations will be discussed in what follows.

We define a transformation matrix which rotates the fixed coordinate
system so that its $z$ axis to be oriented along 
$\vec{\ddot{\mathcal{M}}}_S$. Therefore, all other vector quantities
should be rotated via the defined transformation to be described in
this new rotated coordinate system. If we describe
$\vec{\ddot{\mathcal{M}}}_S$ by polar and azimuthal angles i.e.
$\theta_{\ddot{M}}$ and $\varphi_{\ddot{M}}$, in the fixed original
coordinate system, a rotation around the $\ddot{z}$ axis equal to
$\varphi_{\ddot{M}}$ and then around the rotated $\ddot{y}$ axis
equal to $\theta_{\ddot{M}}$ are required for aligning $\ddot{z}$
axis and $\vec{\ddot{\mathcal{M}}}_S$ orientations. Hence, the
rotation matrices can be respectively given by (see Ref.
\onlinecite{arfken} for more details):

\begin{figure}[t!]
\includegraphics[width=8.5cm,height=4cm]{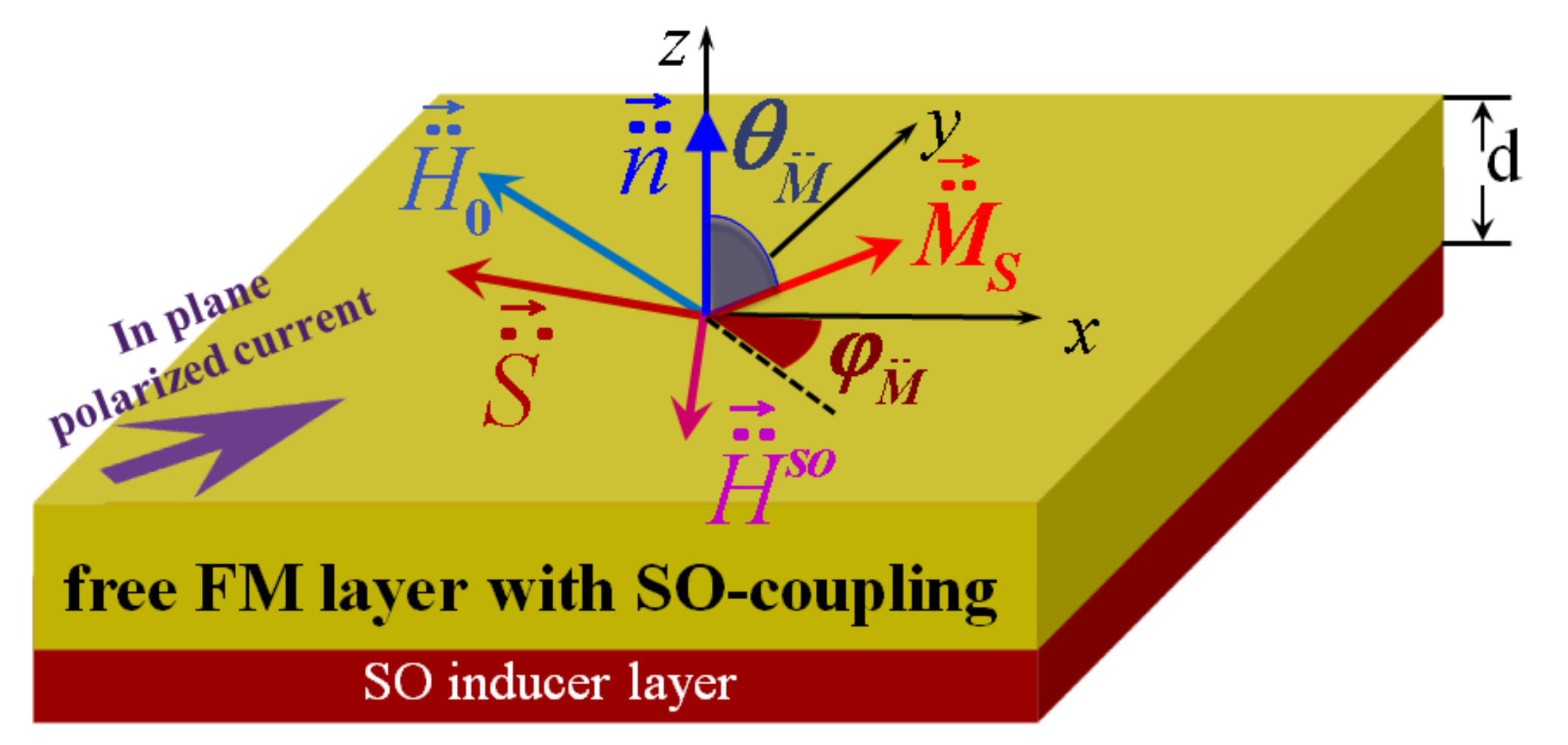}
\caption{\label{fig:model_2} (Color online) Schematic
setup of the free ferromagnetic (FM) layer with a general saturation
magnetization direction, $\vec{\ddot{\mathcal{M}}}_S$, described by polar and azimuthal angles
$\theta_{\ddot{M}}$ and $\varphi_{\ddot{M}}$, respectively. The
thickness of free ferromagnetic layer is denoted by $d$. The
externally applied static magnetic field $\vec{\ddot{H}}_0$,
polarization vector of injected charge current $\vec{\ddot{S}}$,
spin-orbit coupling torque vector $\vec{\ddot{H}}^{so}$, and finally
normal unity vector $\vec{\ddot{n}}$ are shown. The
ferromagnetic film is located in the $\ddot{x}\ddot{y}$ plane so
that $\ddot{z}$ axis is normal to the ferromagnetic film. The
spin-orbit coupling is assumed to be induced via a substrate layer
into the free ferromagnetic layer. The double dot represents the vector
quantities in the non-rotated coordinate system (laboratory
framework). }
\end{figure}

\begin{eqnarray}
&&\nonumber R_z(-\varphi_{\ddot{M}})=\left(\begin{array}{ccc}
                  \cos\varphi_{\ddot{M}} & -\sin\varphi_{\ddot{M}} & 0 \\
                  \sin\varphi_{\ddot{M}} & \cos\varphi_{\ddot{M}} & 0 \\
                  0 & 0 & 1
                \end{array}\right),\\&&\nonumber R_y(\theta_{\ddot{M}})=\left(\begin{array}{ccc}
                  \cos\theta_{\ddot{M}} & 0 & -\sin\theta_{\ddot{M}} \\
                  0 & 1 & 0 \\
                  \sin\theta_{\ddot{M}} & 0 & \cos\theta_{\ddot{M}}
                \end{array}\right).
\end{eqnarray}
The total rotation matrix is thus the multiplication of $R_y$ and
$R_z$ i.e.
\begin{eqnarray}
&&\nonumber R^{t}=R_yR_z=\\&&\left(\begin{array}{ccc}
                  \cos\theta_{\ddot{M}}\cos\varphi_{\ddot{M}} & -\cos\theta_{\ddot{M}}\sin\varphi_{\ddot{M}} & -\sin\theta_{\ddot{M}} \\
                  \sin\varphi_{\ddot{M}} & \cos\varphi_{\ddot{M}} & 0 \\
                  \sin\theta_{\ddot{M}}\cos\varphi_{\ddot{M}} & -\sin\theta_{\ddot{M}}\sin\varphi_{\ddot{M}} &
                  \cos\theta_{\ddot{M}}
                \end{array}\right).
\end{eqnarray}

We characterize each vector quantity by its polar
and azimuthal angle in the fixed original coordinate system shown
in Fig. \ref{fig:model_2}.
Since we assume a homogeneous magnetization texture (macrospin approximation), we have
$\vec{\nabla}^2\vec{\ddot{\mathcal{M}}}=0$.
The total effective field entering the LLG-equation may now be decomposed into the following terms:
\begin{align}
\vec{H}^t &= \vec{H}^{dip}+\vec{h}^{dip}(t)+\vec{H}^{a}+\vec{h}^{a}(t)+\vec{H}^{so}\notag\\
&+b\vec{S}+\vec{H}_0+\vec{h}^{ext}(t)\notag\\
&\equiv\vec{H}+\vec{h}(t).
\end{align}
Above, $\{\vec{H}^{dip}, \vec{h}^\text{dip}(t)\}$ and
$\{\vec{H}^{a},\vec{h}^{a}(t)\}$ are the static and dynamic parts of
the dipole and anisotropy fields respectively, $\vec{H}^{so}$ is the
spin-orbit field, $b\vec{S}$ is the spin-torque effective field
(which is usually negligible), $\vec{H}_0$ is the static externally
applied field, and finally $\vec{h}^{ext}(t)$ is a small rf field
applied perpendicularly to the saturation magnetization direction
$z$ in order to probe the ferromagnetic resonance. To show an
example of how the quantities in the two coordinate systems are
related, note that the $x$, $y$, and $z$ components of the
externally applied static magnetic field $\vec{H}_0$ in the rotated
coordinate system are given by:

\begin{eqnarray}
\nonumber
&&H_{0x}=\ddot{H}_{0}\left\{\cos\theta_{\ddot{M}}\cos\varphi_{\ddot{M}}\sin\theta_{\ddot{H}_0}\cos\varphi_{\ddot{H}_0}-
\right.\\&&\left.\cos\theta_{\ddot{M}}\sin\varphi_{\ddot{M}}\sin\theta_{\ddot{H}_0}\cos\varphi_{\ddot{H}_0}-
\sin\theta_{\ddot{M}}\cos\theta_{\ddot{H}_0}\right\},\\\nonumber
&&H_{0y}=\ddot{H}_{0}\left\{\sin\varphi_{\ddot{M}}\sin\theta_{\ddot{H}_0}\cos\varphi_{\ddot{H}_0}+\right.\\
&&
\left\{\cos\varphi_{\ddot{M}}\sin\theta_{\ddot{H}_0}\cos\varphi_{\ddot{H}_0}\right\},\\\nonumber
&&H_{0z}=\ddot{H}_{0}\left\{\sin\theta_{\ddot{M}}\cos\varphi_{\ddot{M}}\sin\theta_{\ddot{H}_0}\cos\varphi_{\ddot{H}_0}-
\right.\\&&\left.\sin\theta_{\ddot{M}}\sin\varphi_{\ddot{M}}\sin\theta_{\ddot{H}_0}\cos\varphi_{\ddot{H}_0}-
\cos\theta_{\ddot{M}}\cos\theta_{\ddot{H}_0}\right\}.
\end{eqnarray}
As mentioned above, the dipole field can be divided into static
$\vec{H}^{dip}$ and dynamic $\vec{h}^{dip}(t)$ parts. In the rotated
coordinate system they may be obtained as \cite{landeros_prb_08}:
\begin{eqnarray}
&&\nonumber
\vec{H}^{dip}=\mathcal{M}\cos\theta_{\ddot{M}}\left(\begin{array}{c}
                                       \cos\theta_{\ddot{M}}\sin\varphi_{\ddot{M}} \\
                                       -\cos\varphi_{\ddot{M}} \\
                                       \sin\theta_{\ddot{M}}\sin\varphi_{\ddot{M}}
                                     \end{array}
\right),\\&&\nonumber \vec{h}^{dip}(t)=4\pi
m_y(t)\sin\theta_{\ddot{M}}\left(\begin{array}{c}
              \cos\theta_{\ddot{M}}\sin\varphi_{\ddot{M}} \\
              -\cos\varphi_{\ddot{M}} \\
              \sin\theta_{\ddot{M}}\sin\varphi_{\ddot{M}}
            \end{array}
\right),
\end{eqnarray}
where $\mathcal{M}\approx4\pi\mathcal{M}_S-H^{a}$. Assuming a weak rf magnetic field applied transverse to the $\hat{z}$-direction, we may
 consider the components of magnetization in the rotated coordinate
 system as
$\mathcal{M}_z=\mathcal{M}_S\gg\mathcal{M}_x, \mathcal{M}_y$. In
this case, the following time-dependent coupled differential
equations for the precessing magnetization components are obtained;

\begin{eqnarray}
  &&\nonumber\frac{\partial \mathcal{M}_x}{\partial t}= -\gamma \mathcal{M}_y H_z^t+\gamma \mathcal{M}_x(\beta H_{z}^{so}+P_{s}S_{z})\\&&
  +\gamma\mathcal{M}_S(H_y^t-(\beta H_{x}^{so}+P_{s}S_{x}))-\alpha \frac{\partial \mathcal{M}_y}{\partial t}, \nonumber
  \\&&\nonumber\frac{\partial \mathcal{M}_y}{\partial t}= \gamma \mathcal{M}_x H_z^t+\gamma \mathcal{M}_y(\beta H_{z}^{so}+P_{s}S_{z})\\&&
  -\gamma\mathcal{M}_S(H_x^t+(\beta H_{y}^{so}+P_{s}S_{y}))+\alpha \frac{\partial \mathcal{M}_x}{\partial t}, \nonumber
  \\&&\nonumber\frac{\partial \mathcal{M}_z}{\partial t}=\frac{\partial \mathcal{M}_S}{\partial t}=0=\gamma\mathcal{M}_x(-H_y^t+(\beta H_{x}^{so}+P_{s}S_{x}))
  \\&&+\gamma\mathcal{M}_y(H_x^t+(\beta
  H_{y}^{so}+P_{s}S_{y})). \nonumber
   \end{eqnarray}
Setting the transverse part of the magnetization and fields equal to
zero in the above equations for $\partial_t\mathcal{M}_x$ and
$\partial_t\mathcal{M}_y$, one obtains the equilibrium conditions
which specify the orientation of the $z$ axis:
\begin{eqnarray}\label{eq:condition}
\left\{\begin{array}{c}
  H_x+(\beta_{so}H_{y}^{so}+\beta_{s}S_{y}) = 0 \\
  H_y-(\beta_{so}H_{x}^{so}+\beta_{s}S_{x}) = 0 \\
\end{array}\right..
 \end{eqnarray}
This is consistent with the equation for $\partial_t\mathcal{M}_z$
and our preassumption namely;
$\mathcal{M}_z\gg\mathcal{M}_x,\mathcal{M}_y$ . In order to obtain
the solution for the transverse components $\mathcal{M}_x$ and
$\mathcal{M}_y$ to lowest order, we now substitute these conditions
back into the equations of motion for the magnetization components
above and obtain:
\begin{eqnarray}\label{llg_3}
  \nonumber\frac{\partial \mathcal{M}_x}{\partial t}&=& -\gamma \mathcal{M}_y H_z+\gamma \mathcal{M}_Sh_y(t)-\alpha \frac{\partial \mathcal{M}_y}{\partial t} \nonumber
  \\&&\nonumber+\gamma\mathcal{M}_x(\beta H^{so}_z+P_{s}S_z),\\
  \frac{\partial \mathcal{M}_y}{\partial t}&=& +\gamma \mathcal{M}_x H_z-\gamma \mathcal{M}_Sh_x(t)+\alpha
  \frac{\partial \mathcal{M}_x}{\partial t}\\&&\nonumber+\gamma\mathcal{M}_y(\beta
  H^{so}_z+P_{s}S_z).
\end{eqnarray}
In our calculations we have set the time-dependent fields sufficienty small so that those terms including higher orders of
time-dependent components are negligible. Assuming that the
the external time-dependent magnetic field induces the same
frequency in all time-dependent components of other vector
quantities (including responses) as itself, $\Omega$, we get \eg
$\vec{h}^{dip}(t)=\vec{h}^{dip}e^{-i\Omega t}$.
By substituting this time-dependency into Eqs. (\ref{llg_3}) we
arrive at $\tilde{\mathcal{M}}(t)=\chi \tilde{h}^{ext}(t)$ in which
$\tilde{\mathcal{M}}(t)=(\mathcal{M}_x,\mathcal{M}_y)^T$,
$\tilde{h}^{ext}(t)=(h_x^{ext},h_y^{ext})^T$, and;
\begin{eqnarray}
\chi=\left(
          \begin{array}{cc}
            \chi_{xx} & \chi_{xy} \\
            \chi_{yx} & \chi_{yy} \\
          \end{array}
        \right).
\end{eqnarray}
$\chi$ is known as the susceptibility tensor which determines the
behavior of magnetization in response to the external time-dependent
magnetic field. The components of the obtained susceptibility tensor
in the presence of spin-orbit coupling read:
\begin{eqnarray}
  \nonumber\chi_{xx} &=& +\Gamma\left\{\gamma \mathcal{W}_y\Xi-\Delta\alpha\Omega-i(\gamma\Delta \mathcal{W}_y+\Omega\alpha\Xi)\right\}, \\
  \nonumber\chi_{xy} &=& -\Gamma\left\{\Sigma\Xi+\Delta\Omega-i(\Delta\Sigma-\Omega\Xi)\right\}, \\
  \nonumber\chi_{yx} &=& +\Gamma \left\{\Sigma\Xi+\Delta\Omega-i(\Delta\Sigma-\Omega\Xi)\right\}\\
  \nonumber\chi_{yy} &=& +\Gamma\left\{\gamma \mathcal{W}_x\Xi-\Delta\alpha\Omega-i(\gamma\Delta
  \mathcal{W}_x+\Omega\alpha\Xi)\right\},
\end{eqnarray}
where we have defined the following parameters;
\begin{eqnarray}\label{param}
    \nonumber&&\Gamma=\frac{\gamma
\mathcal{M}_S}{\Xi^2+\Delta^2},\;\;\;\;\Sigma=\gamma
    (\beta_{so}H^{so}_z+\beta_{s}S_z),\\&&\nonumber
    \Xi=\Upsilon^2-\Omega^2(1+\alpha^2),\;\;\;\;\Upsilon=\sqrt{\gamma^2\mathcal{W}_x\mathcal{W}_y+\Sigma^2},\\\nonumber
    &&\Delta=2\Sigma\Omega-\gamma\alpha\Omega(\mathcal{W}_x+\mathcal{W}_y),\\&&\nonumber
\mathcal{W}_x=H_z+\mathcal{M}\sin\theta_{\ddot{M}}\cos\theta_{\ddot{M}}\sin\varphi_{\ddot{M}},
\\\nonumber
&&\mathcal{W}_y=H_z+\mathcal{M}\sin\theta_{\ddot{M}}\cos\varphi_{\ddot{M}}.
\end{eqnarray}
The susceptibility tensor components may be used to compute
physical quantities of interest such as the absorbed power (which
is experimentally relevant \cite{prb_86_dreher}) by the
ferromagnetic sample with volume $V$ at frequency $\Omega$. In turn, this
gives a clear signal of ferromagnetic resonance in the
absorption spectrum. This energy dissipation is given by
$P_{power}^{abs}=$Im$\{P_{power}\}$ where $P_{power}$ is defined by:
\begin{eqnarray}
&&\nonumber P_{power}=-\frac{\Omega}{2}\int_V
dV\vec{h}^{{ext*}}\cdot\vec{\mathcal{M}}=-\frac{\Omega}{2}\int_V
dV\vec{h}^{{ext*}}\cdot\chi\vec{h}^{ext}\\&&\nonumber=-\frac{\Omega}{2}\int_V
dV
\left\{{|h_x^{ext}|}^2\chi_{xx}+h_x^{ext}*h_y^{ext}\chi_{xy}+\right.\\&&\nonumber
\left.
h_y^{ext}*h_x^{ext}\chi_{yx}+{|h_y^{ext}|}^2\chi_{yy}\right\}.
\end{eqnarray}
This expression simplifies if the rf magnetic field only has one
component, \eg $\vec{h}(t)=h_x^{ext}(t)$, in which case the power
absorbed at radio-frequency $\Omega$ can be expressed by:
\begin{eqnarray}
\nonumber P_{power}^{abs} =\frac{\Omega}{2}\int_V dV \frac{\gamma
\mathcal{M}_S {|h_x^{ext}|}^2}{\Xi^2+\Delta^2}(\gamma\Delta
\mathcal{W}_y+\Omega\alpha\Xi).
\end{eqnarray}
Although the above expressions may be numerically evaluated in our system for a specific parameter choice, we focus below on analytical insights that may be gained. In particular, we are interested
in the role played by spin-orbit interactions and the magnitude/direction of the injected current. So far, our treatment has been
general and accounted for several terms contributing to the
susceptibility tensor. In order to identify the role
played by current-dependent spin-orbit coupling in the ferromagnetic
resonance, we need to derive an analytical expression
for the ferromagnetic resonance frequency $\Omega_\text{FMR}$. This
is defined as the frequency where the $P_{power}^{abs}$ has a
maximum. In their general form shown above, this cannot be done
analytically in an exact manner. However, progress can be made by
considering the denominator of $P_{power}^{abs}$. This quantity has the
following form when all the frequency-dependence is written
explicitly:
\begin{align}\label{eq:gamma}
\Xi^2+\Delta^2 &= [\Upsilon^2 -
\Omega^2(1+\alpha^2)]^2 \notag\\
&+ \Omega^2[2\Sigma - \gamma\alpha(\mathcal{W}_x+\mathcal{W}_y)]^2.
\end{align}
Following the standard procedure of neglecting the second term
above, one may identify the resonance frequency similarly to Ref.
\onlinecite{landeros_prb_08} as $\Omega_\text{FMR} = \Upsilon$. We have also verified that this holds numerically for a realistic parameter set.

To see how the spin-orbit coupling affects $\Omega_\text{FMR}$, one
should note in particular its dependence on the current $J$. It is
instructive to consider first the scenario with zero spin-orbit
coupling, in which case the resonance frequency may be written as:
\begin{align}\label{eq:fmrsimple}
\Omega_\text{FMR} = \sqrt{c_1 + c_2 J^2},
\end{align}
where $c_1$ and $c_2$ are determined by the quantities in Eq.
(\ref{param}) in the limit $\tilde{\alpha}_R \to 0$. Importantly,
they are independent on the current bias $J$, which means that the
resonance frequency is completely independent on the
\textit{direction} of the applied current as it is only the
magnitude $J^2$ that enters. Therefore, the current direction cannot
alter the $\Omega_\text{FMR}$. Turning on the spin-orbit coupling so that
$\tilde{\alpha}_R \neq 0$, one may in a similar way show from the
above equations that the resonance frequency now can be written as:
\begin{align}\label{eq:omegasoc}
\Omega_\text{FMR} = \sqrt{(d_1+\mathcal{D}J)(d_2+\mathcal{D}J) + d_3J^2},
\end{align}
where again the coefficients $d_i$ and $\mathcal{D}$ are determined
from Eq. (\ref{param}). It then follows from Eq. (\ref{eq:omegasoc})
that the resonance frequency will be \textit{asymmetric} with
respect to the applied current direction when spin-orbit coupling is
present. In particular, one obtains different values for
$\Omega_\text{FMR}$ by reversing the current $J\to(-J)$ so that the
$\mathbb{Z}_2$ symmetry in Eq. (\ref{eq:fmrsimple}) is lost. The
main signature of spin-orbit coupling in the current-biased
ferromagnetic resonance setup under consideration is then an
asymmetric current dependence which should be distinguishable from
the scenario without spin-orbit interactions. It is interesting to
note that the current-dependence on the ferromagnetic resonacne and
the linewidth allows one to exert some control over the
magnetization dissipation/absorption in the system via $J$. The
presence of spin-orbit interactions enhances this control since it
introduces a directional dependence which is absent without such
interactions.

\section{Summary}\label{sec:summary}

In summary, we have considered the influence of existence of
spin-orbit interactions on both domain wall motion and ferromagnetic
resonance of a ferromagnetic film. Due to the coupling between the
momentum and spin of the electrons, the degeneracy between domain
wall textures is broken which in turn leads to qualitatively
different behavior for various wall profiles, \eg Bloch vs. Neel
domain walls. By taking into account both the field- and
Slonczewski-like spin-orbit torque, we have derived exact analytical
expressions for the wall velocity and the onset of Walker breakdown.
One of the most interesting consequences of the spin-orbit torques
is that they render Walker breakdown to be universal for some wall
profiles in the sense that the threshold is completely independent
on the material-dependent damping $\alpha$, non-adiabaticity
$\beta$, and the chirality $\sigma$ of the domain wall. We have also
shown that domain wall motion against the current flow is sustained
in the presence of multiple spin-orbit torques and that the wall
profile will determine the qualitative influence of these different
types of torques. Finally, we calculated the ferromagnetic resonance
response of a ferromagnetic material in the presence of spin-orbit
torques, i.e. a setup with a current bias. We found a key signature
of the spin-orbit interactions in the resonance frequency, namely
that the latter becomes asymmetric with respect to the direction of
current injection. This is different from usual ferromagnets in the
presence of spin-transfer torques in the absence of spin-orbit
interactions, where the frequency is found to be symmetric with
respect to the current direction.

\acknowledgments

M.A. would like to thank P. Landeros for helpful conversations and
A. Sudb{\o} for valuable discussion.

\end{document}